\documentclass[fleqn,usenatbib]{mnras}

\usepackage[T1]{fontenc}

\DeclareRobustCommand{\VAN}[3]{#2}
\let\VANthebibliography\thebibliography
\def\thebibliography{\DeclareRobustCommand{\VAN}[3]{##3}\VANthebibliography}

\usepackage{graphicx}	
\usepackage{amsmath}	
\usepackage{amssymb}	
\usepackage{color}
\usepackage{amsmath}
\usepackage{threeparttable}
\usepackage{newtxtext,newtxmath}

\newcommand{\ie}{\textit{i.e.,}}
\newcommand{\eg}{\textit{e.g.,}}

\newcommand{\logM}{$\log(M/\rm{M}_\odot)$}
\newcommand{\zphot}{$z_\textrm{phot}$}

\newcommand{\OII}{\hbox{{\rm [O}\kern 0.1em{\sc ii}{\rm ]$\lambda\lambda3726,3729$}}}
\newcommand{\OIII}{\hbox{{\rm [O}\kern 0.1em{\sc iii}{\rm ]$\lambda\lambda4959,5007$}}}

\newcommand{\PSC}{\mbox{Elent\'ari}}

\title[\PSC: A $z\sim3.3$ Proto-Supercluster in COSMOS]{\sc \PSC: A massive proto-supercluster at $z\sim3.3$ in the COSMOS field}
\author[B. Forrest et al.]{
Ben Forrest,$^{1, 2}$\thanks{E-mail: bforrest@ucdavis.edu}
Brian C. Lemaux,$^{1, 3}$
Ekta Shah,$^{1}$
Priti Staab,$^{1}$
Ian McConachie,$^{2}$
\newauthor{
Olga Cucciati,$^{4}$
Roy R. Gal,$^{5}$
Denise Hung,$^{5}$
Lori M. Lubin,$^{1}$
Letizia P. Cassar\`a,$^{6}$
}
\newauthor{
Paolo Cassata,$^{7, 8}$
Wenjun Chang,$^{2}$
M.C. Cooper,$^{9}$
Roberto Decarli,$^{4}$
Percy Gomez,$^{10}$
}
\newauthor{
Gayathri Gururajan,$^{11, 4}$
Nimish Hathi,$^{12}$
Daichi Kashino,$^{13}$
Danilo Marchesini,$^{14}$
}
\newauthor{
Z. Cemile Marsan,$^{15}$
Michael McDonald,$^{2}$
Adam Muzzin,$^{15}$
Lu Shen,$^{16, 17}$
}
\newauthor{
Stephanie Urbano Stawinski,$^{9}$
Margherita Talia,$^{11, 4}$
Daniela Vergani,$^{4}$
Gillian Wilson,$^{2, 18}$
}
\newauthor{
Giovanni Zamorani$^{4}$
}
\\
\\
$^{1}$ Department of Physics and Astronomy, University of California Davis, One Shields Avenue, Davis, CA, 95616, USA\\
$^{2}$ Department of Physics \& Astronomy, University of California Riverside, 900 University Ave., Riverside, CA, 92521, USA \\
$^{3}$ Gemini Observatory, NSF's NOIRLab, 670 N. A'ohoku Place, Hilo, HI, 96720, USA\\
$^{4}$ INAF Osservatorio di Astrofisica e Scienza dello Spazio di Bologna, Via Piero Gobetti 93/3, 40129 Bologna, Italy\\
$^{5}$ University of Hawai'i, Institute for Astronomy, 2680 Woodlawn Drive, Honolulu, HI 96822, USA\\
$^{6}$ INAF-IASF Milano, Via Alfonso Corti 12, 20133, Milano, Italy \\
$^{7}$ Dipartimento di Fisica e Astronomia Galileo Galilei, Universit\`a degli Studi di Padova, Vicolo dell'Osservatorio 3, 35122 Padova, Italy\\
$^{8}$ INAF Osservatorio Astronomico di Padova, Vicolo dell'Osservatorio 5, 35122 Padova, Italy\\
$^{9}$ Department of Physics and Astronomy, University of California, Irvine, 4129 Frederick Reines Hall, Irvine, CA 92697, USA \\
$^{10}$ W.M. Keck Observatory, 65-1120 Mamalahoa Hwy., Kamuela, Hawai'i 96743, USA \\
$^{11}$ University of Bologna - Department of Physics and Astronomy ``Augusto Righi'' (DIFA), Via Gobetti 93/2, I-40129, Bologna, Italy\\
$^{12}$ Space Telescope Science Institute, Baltimore, MD 21218, USA\\
$^{13}$ Institute for Advanced Research, Nagoya University, Furocho, Chikusa-ku, Nagoya, 464-8601, Japan\\
$^{14}$ Department of Physics and Astronomy, Tufts University, 574 Boston Avenue, Medford, MA 02155, USA \\
$^{15}$ Department of Physics and Astronomy, York University, 4700, Keele Street, Toronto, ON MJ3 1P3, Canada \\
$^{16}$ Department of Physics and Astronomy, Texas A\&M University, College Station, TX, 77843-4242 USA\\
$^{17}$ George P.\ and Cynthia Woods Mitchell Institute for Fundamental Physics and Astronomy, Texas A\&M University, College Station, TX, 77843-4242 USA\\
$^{18}$ Department of Physics, University of California, 5200 North Lake Rd., Merced, CA 95343, USA\\
}

\date{Accepted XXX. Received YYY; in original form ZZZ}

\pubyear{2023}

\begin{document}
\label{firstpage}
\pagerange{\pageref{firstpage}--\pageref{lastpage}}
\maketitle

\begin{abstract}

Motivated by spectroscopic confirmation of three overdense regions in the COSMOS field at $z\sim3.35$, we analyze the uniquely deep multi-wavelength photometry and extensive spectroscopy available in the field to identify any further related structure.
We construct a three dimensional density map using the Voronoi tesselation Monte Carlo method and find additional regions of significant overdensity.
Here we present and examine a set of six overdense structures at $3.20<z<3.45$ in the COSMOS field, the most well characterized of which, PCl~J0959+0235, has 80 spectroscopically confirmed members and an estimated mass
of $1.35\times 10^{15}$~M$_\odot$, and is modeled to virialize at $z\sim1.5-2.0$.
These structures contain ten overdense peaks with $>5\sigma$ overdensity separated by up to 70 cMpc, suggestive of a proto-supercluster similar to the Hyperion system at $z\sim2.45$.
Upcoming photometric surveys with \textit{JWST} such as COSMOS-Web, and further spectroscopic follow-up will enable more extensive analysis of the evolutionary effects that such an environment may have on its component galaxies at these early times.
\end{abstract}




\section{Introduction}

Galaxy clusters are dense neighborhoods of galaxies whose members are gravitationally bound, with halo masses of \mbox{log($M$/M$_\odot)\gtrsim14$} \citep[\eg][]{vanderBurg2014,Balogh2020,Hung2021}.
Associations of galaxy clusters, known as superclusters, are, while rare, the largest overdense structures in the Universe \citep[\eg][]{Raychaudhury1991, Tully2014, Einasto2021}, though many of them may not in fact gravitationally collapse in the future \citep{Chon2015, Remus2023}.
Regardless, the progenitors of such structures should in principle be detectable with significant amounts of spectroscopic data and high quality photometry over substantial fields-of-view.

The progenitors of galaxy clusters, protoclusters, are predicted to extend over $\sim10-20$~cMpc at $z\sim2$ and even larger areas at earlier epochs \citep[\eg][]{Chiang2013}.
Indeed, such systems have been identified out to high redshifts \citep[\eg][]{Steidel1998, Venemans2002, Ouchi2005, Capak2011, Ito2023}, though in many cases in-depth characterizations of the systems are not possible without further spectroscopic information.
Extensive follow-up of several protocluster systems have unveiled much larger, extended structures with multiple overdensity peaks, as in SSA22 \citep[\eg][]{Steidel1998, Yamada2012}, the Spiderweb \citep[\eg][]{Pentericci2000, Jin2021}, Hyperion \citep[][and references therein]{Cucciati2018}, PCl~J0227-0421 \citep{Lemaux2014, Shen2021}, PCl~J1001+0220 \citep[][P. Staab, in prep]{Lemaux2018}, and PCl~J0332-2749 \citep[][E. Shah, in preparation]{Forrest2017}.

\begin{figure*}
    \includegraphics[width=\textwidth, trim=0in 6in 0in 0in]{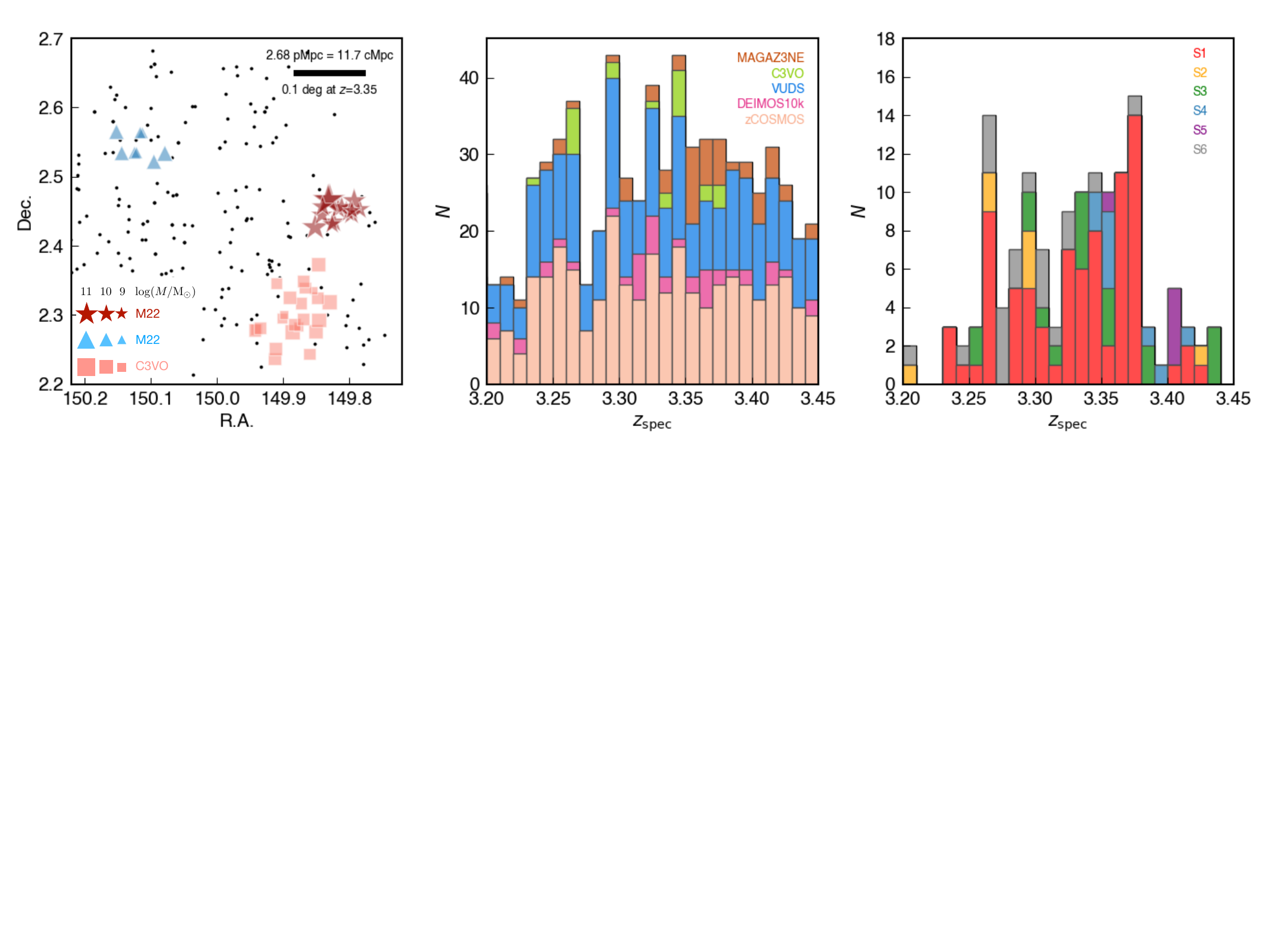}
    \caption{Spectroscopically confirmed galaxies at $3.20<z_{\rm spec}<3.45$ in the COSMOS field. \textbf{Left:} Zoom-in of the portion of the field targeted independently by MAGAZ3NE and C3VO. Dark red stars and blue triangles are MOSFIRE spectroscopic redshifts published in \citet{McConachie2022}, while salmon squares are MOSFIRE spectroscopic redshifts followed up in C3VO. These shapes are sized by the logarithm of stellar mass. Black dots are other spectroscopically confirmed galaxies from zCOSMOS, DEIMOS10k, and VUDS at $3.20<z<3.45$. \textbf{Center:} Stacked histogram of spectroscopically confirmed galaxies in the range $3.20<z_{\rm spec}<3.45$ over the entire COSMOS field from the listed surveys. \textbf{Right:} Stacked histrogram of spectroscopically confirmed members in each of the six structures discussed in this work.}
    \label{fig:Specz}
\end{figure*}

Recently, the Massive Ancient Galaxies at $z>3$ Near-infrared survey \citep[MAGAZ3NE;][]{Forrest2020b} detected two protoclusters with 22 spectroscopically-confirmed members at $z\sim3.37$ separated by 35~cMpc \citep{McConachie2022} in the COSMOS field.
Independently, the Charting Cluster Construction with VUDS and ORELSE survey \citep[C3VO;][]{Shen2021, Lemaux2022} followed up another nearby candidate overdensity and confirmed an additional 22 galaxies at $3.26<z_{\rm spec}<3.38$ only 24 cMpc from the MAGAZ3NE structure (Figure~\ref{fig:Specz}).
The close spatial proximity of so many spectroscopically confirmed galaxies is suggestive of a much larger system similar to the Hyperion proto-supercluster and motivated a further analysis of available data in the COSMOS field to search for evidence of such structure.

We present here the results of this search which includes six extended overdense structures at $z\sim3.3$ in the COSMOS field and describe their spatial extent through a combination of deep spectroscopic and photometric analysis.
The entire system is identified as PCl~J0959+0235, and we refer to it with the nickname \PSC, due to the large number of new stars formed in such protoclusters.
The observational data used for this study are described in Section~\ref{Sec:Data}, the process of structure identification in Section~\ref{Sec:ID}, and potential evolution of these systems in Section~\ref{Sec:Evo} before concluding.
A companion paper, B. Forrest et. al., in preparation, will provide a more in-depth discussion of methods used.
Throughout this work we assume a \citet{Chabrier2003} IMF, AB magnitude system \citep{Oke1983}, and a cosmology with \mbox{$H_0=70$ km/s/Mpc}, \mbox{$\Omega_M=0.3$}, and \mbox{$\Omega_\Lambda=0.7$}.

\section{Data} \label{Sec:Data}

The COSMOS field \citep{Scoville2007, Koekemoer2007} has a long history of extensive observations, enabling the construction of deep, multi-band photometric catalogs, including those from \citet{Capak2007}, \citet{Ilbert2009}, \citet[][UltraVISTA DR1]{Muzzin2013a},  \citet[][COSMOS2015]{Laigle2016}, \citet{Nayyeri2017}, and \citet[][COSMOS2020]{Weaver2022a}.
These photometric catalogs and their associated derived properties, including well-characterized photometric redshifts (\zphot), rest-frame colors, stellar masses, and star formation rates (SFRs) have been the basis for numerous spectroscopic campaigns.

\subsection{Spectroscopy}\label{sec:spec}

In this work we make use of spectral data from several spectroscopic surveys targeting the COSMOS field including the zCOSMOS survey \citep{Lilly2007}, the VIMOS Ultra-Deep Survey \citep[VUDS; ][]{LeFevre2015}, the DEIMOS 10k Spectroscopic Survey \citep{Hasinger2018}, the Massive Ancient Galaxies at $z>3$ Near-Inrared survey \citep[MAGAZ3NE;][]{Forrest2020b}, and the Charting Cluster Construction with VUDS and ORELSE survey \citep[C3VO;][]{Lemaux2022}.
Together, these surveys have targeted over 40000 individual galaxies in the COSMOS field using the VLT/VIMOS \citep{Lefevre2003}, Keck/DEIMOS \citep{Faber2003}, and Keck/MOSFIRE \citep{McLean2010, McLean2012} instruments.

For information on the data reduction process for zCOSMOS \citep{Lilly2007}, VUDS \citep{LeFevre2015}, and DEIMOS 10k \citep{Hasinger2018} data, we refer the reader to the associated papers.
We use data from both zCOSMOS-Bright and zCOSMOS-Deep, including an updated zCOSMOS-Deep catalog (D. Kashino, priv. comm.).
MAGAZ3NE MOSFIRE spectra were (re-)reduced using the MOSDEF 2D data reduction pipeline \citep{Kriek2015}, and MOSFIRE data from the C3VO survey were reduced similarly.
A custom Python code was used to perform optimal spectral extraction of the targets \citep{Horne1986}.
When one or more emission lines were present, the shape of the optimal aperture was calculated around the strongest emission line.
When no trace was identified, a boxcar extraction was used at the intended location on the mask.

Broadly speaking, zCOSMOS, VUDS, and DEIMOS10k observed galaxies across the field based on cuts in optical magnitude, color, and photometric redshift.
These surveys use spectrographs which target the rest-frame ultraviolet at the redshifts considered here resulting in a higher likelihood of confirming star-forming galaxies than quiescent galaxies.
Additionally, given the survey and instrument designs, regions with large projected overdensities of galaxies can have lower spectroscopic sampling rates than the field.
In contrast, the C3VO and MAGAZ3NE surveys chose specific regions to target based on candidate overdensities and ultra-massive galaxies, respectively.
While still more likely to confirm emission line galaxies, these surveys have more complex selection functions and increase the sampling rate in overdense regions.
The resulting uneven sampling bias which must be considered when analyzing results, and the effects are discussed briefly in section 3.2.
A more detailed investigation of this issue will be provided in a follow-up paper.

\subsection{Catalog Matching}

While target selection for the various spectroscopic surveys used in this work came from different catalogs, we compare the spectroscopic redshifts, sky coordinates, and $i$- and $K$-band magnitudes from the spectroscopic survey selection catalog to the photometric redshifts \citep[\texttt{LePhare};][]{Arnouts1999, Ilbert2006}, coordinates and magnitudes in the same bandpasses from the COSMOS2020 Classic photometric catalog \citep{Weaver2022a} to find the best-match photometric galaxy to each spectroscopically-confirmed galaxy.

In total we return over 40000 objects which were spectroscopically targeted by at least one survey and were given quality flags consistent with the flagging system from VUDS \citep{LeFevre2015}.
Nearly 20000 of these have high quality spectroscopic reliability flags = 3 or 4, which indicate a $>95\%$ or 100\% confidence in the spectroscopic redshift, respectively.
For galaxies in this work, such flags indicate the high signal-to-noise detection of multiple emission lines.
This includes $\sim900$ objects with spectroscopic redshifts which do not have a COSMOS2020 object satisfying our matching criteria.

\begin{figure*}
	\includegraphics[width=\textwidth, trim=0in 0.5in 0in 0in]{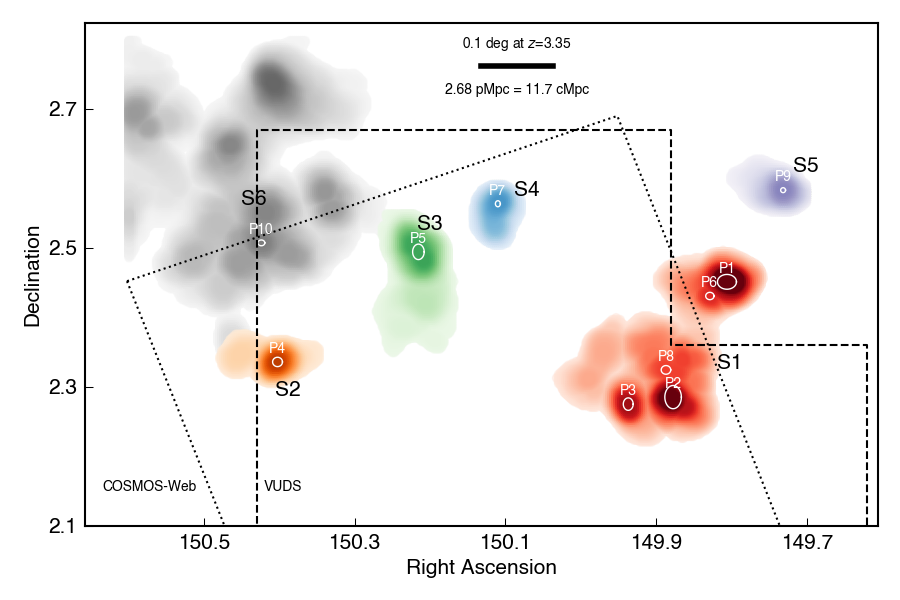}
    \caption{Projected overdensity map of \PSC. The ten overdense peaks (P\#; $\sigma_\delta>5$) with $3.20<z_{\rm peak}<3.45$ and total mass \logM$>12$ are shown as white ellipses (with axes $R_x, R_y$ as in Table~\ref{tab:peaks_id}) and the six overdense structures (S\#; $\sigma_\delta>2$) in which they reside are colored shapes, with darker shades indicating more overdense regions. The black dashed and dotted lines show the footprints of VUDS and COSMOS-Web \citep{Casey2022}, respectively.  The dark red stars and salmon squares from Figure~\ref{fig:Specz} lie in S1, while the blue triangles lie in S4. Notably, S1 (red) contains five of the ten overdense peaks, while S6 appears quite extended possibly due in part to the reduced spectroscopic coverage outside of the VUDS footprint and reaches the bounds of the VMC map. }
    \label{fig:Structure}
\end{figure*}

\begin{table*}
	\caption{The properties of $>5\sigma$ overdense peaks identified between $3.20<z<3.45$ with log($M_{\rm tot}$/M$_\odot$)>12 ordered by total mass. Listed properties include the barycentric center of each peak (2-4), the average overdensity (5), effective radius in each dimension (6-8), observed elongation along the line of sight (9), observed volume (10), total mass (11), the volume and average overdensity corrected for observed elongation (12-13), and the structure in which each peak lies (14). See Section~\ref{Sec:PChar}, \citet{Cucciati2018} and \citet{Shen2021} for more details.}
	\begin{center}
	\begin{threeparttable}
	\begin{tabular}{cccccccccccccc}
		\hline
		ID	& R.A.$_{\rm peak}$	& Dec.$_{\rm peak}$	& $z_{\rm peak}$	& $\langle \delta_{\rm gal} \rangle$	& $R_x$	& $R_y$	& $R_z$	& $E_{z/xy}$	& $V$		& $M_{\rm tot}$		& $V_{\rm corr}$	& $\langle \delta_{\rm gal} \rangle_{\rm corr}$ & Structure\\
			& (deg)			& (deg)			&				& 
				& (cMpc)	& (cMpc)	& (cMpc)	&		& (cMpc$^3$)	& ($10^{14}M_\odot$) &  (cMpc$^3$) & & \\
		(1) & (2) & (3) & (4)  & (5)  & (6) & (7) & (8)  & (9)  & (10) & (11) & (12)  &  (13) & (14) \\
		\hline
                 P1 & 149.8060 & 2.4510 & 3.366 &  3.67 & 1.47 & 1.24 & 8.37 & 6.18 & 851.7 & 0.76 & 137.8 & 12.49 & S1\\
                 P2 & 149.8774 & 2.2850 & 3.341 &  2.87 & 1.28 & 1.94 & 9.07 & 5.64 & 708.9 & 0.56 & 125.8 & 9.85 & S1\\
                 P3 & 149.9369 & 2.2749 & 3.269 &  2.76 & 0.75 & 1.03 & 8.12 & 9.08 & 304.7 & 0.23 & 33.5 & 16.17 & S1 \\
                 P4 & 150.4021 & 2.3356 & 3.255 &  2.54 & 0.75 & 0.77 & 5.61 & 7.35 & 163.5 & 0.12 & 22.2 & 12.37 & S2 \\
                 P5 & 150.2153 & 2.4941 & 3.343 &  2.25 & 0.89 & 1.29 & 3.62 & 3.31 & 151.8 & 0.11 & 45.8 & 4.72 & S3 \\
                 P6 & 149.8287 & 2.4306 & 3.315 &  2.34 & 0.64 & 0.61 & 3.54 & 5.66 & 62.9 & 0.05 & 11.1 & 8.92 & S1 \\
                 P7 & 150.1099 & 2.5634 & 3.354 &  2.16 & 0.38 & 0.50 & 6.41 & 14.45 & 34.6 & 0.02 & 2.4 & 23.52 & S4 \\
                 P8 & 149.8869 & 2.3242 & 3.248 &  2.16 & 0.73 & 0.72 & 1.11 & 1.54 & 25.7 & 0.02 & 16.7 & 1.62 & S1 \\
                 P9 & 149.7315 & 2.5829 & 3.408 &  2.16 & 0.38 & 0.41 & 5.41 & 13.63 & 22.8 & 0.02 & 1.7 & 22.14 & S5 \\
                 P10 & 150.4236 & 2.5071 & 3.409 & 2.17 & 0.54 & 0.49 & 1.52 & 2.94 & 18.1 & 0.01 & 6.1 & 4.00 & S6
	\end{tabular}
	\end{threeparttable}
	\end{center}
        \label{tab:peaks_id}
\end{table*}

\section{Identification of Structure}\label{Sec:ID}

\subsection{Voronoi Monte Carlo Mapping}

We use Voronoi tessellation Monte Carlo (VMC) mapping to quantify the environmental density of the entire COSMOS field at $3.0<z<3.7$.
This process constructs Voronoi cells in projection over narrow redshift windows including spectroscopic redshifts and random draws from photometric redshift probability distributions, which are then averaged over many iterations and resampled onto a fine grid of 3D cells to estimate density.
This methodology has been used and tested extensively in previous works \citep{Lemaux2017, Tomczak2017, Lemaux2018, Cucciati2018, Hung2020, Shen2021} to which we refer the interested reader for further details, and the method used here is identical to that presented in \citet{Lemaux2022}.

From this density mapping, which takes into consideration both photometric and spectroscopic redshift information for catalog objects, we calculate the overdensity, $\delta_{\rm gal}$ in each cell.
In order to account for both density changes as a function of redshift and the potential of biases from large projected under- or overdensities, we then calculate the mean and standard deviation of a gaussian fit to the projected density in redshift slices 7.5~pMpc in depth from $3.0<z<3.7$ over the entire field and fit a $5^{th}$-order polynomial to these data as a function of redshift. 
These fit values to the density distribution are then used to calculate the significance of the overdensity in each voxel, $\sigma_\delta$, which we use to identify extended overdense regions.

\subsection{Peak Identification and Characterization}\label{Sec:PChar}

As in \citet{Cucciati2018} and \citet{Shen2021}, we begin by considering all voxels with an overdensity significance of $\sigma_\delta>5$ in order to identify the peaks of overdense regions.
Such a peak is identified by finding the voxel with the highest significance overdensity and searching for all contiguous voxels with $\sigma_\delta>5$.
All voxels in this peak are removed from the sample and the process is repeated until no remaining voxels have $\sigma_\delta>5$.
Each identified peak then has a total mass  $M_{\rm tot} = \rho_m V (1+\delta_{\rm m})$, where $\rho_m$ is the comoving matter density, $V$ is the volume of the $\sigma_\delta>5$ envelope, \ie\ the sum of the volume of each voxel in the peak, and $\delta_{\rm m}$ is the mass overdensity in the peak.
The mass overdensity is a scaling of the average galaxy overdensity in the peak, $\delta_{\rm gal}$, by a bias factor.
In this work we adopt a bias factor $b=3.1$, as found in \citet{Durkalec2015} at $z\sim3.3$ in the VUDS sample.
The barycentric position, \mbox{$X_{\rm peak} = \Sigma_i ( \delta_{\rm gal, X_i} X_i ) / \Sigma_i (\delta_{\rm gal, X_i})$} for $X=\{$R.A., Dec., $z\}$, and effective radius, \mbox{$R_X= \sqrt{ \Sigma_i (  \delta_{\rm gal, X_i}  (X_i-X_{\rm peak} )^2 ) / \Sigma_i \delta_{\rm gal, X_i}}$}, of each peak are also calculated as in \citet{Cucciati2018, Shen2021}.
This method identifies 10 peaks with $3.20<z_{\rm peak}<3.45$ and total masses log($M_{\rm tot}$/M$_\odot$)>12 in our maps, all of which we consider in this work.

Due to photometric redshift uncertainties and peculiar velocities, the measured effective radius in the redshift dimension is typically elongated with respect to the transverse dimensions.
To account for this effect we make the simplistic assumption that the size of the peak in the redshift dimension should be similar to that in the two projected dimensions \citep[see, \eg][]{Cucciati2018}, and thus calculate a correction factor ($E_{z/xy}$) to correct the volume of the peak for the observed elongation, $V_{\rm corr} = V/E_{z/xy}$.
Similarly, we compute a corrected average galaxy overdensity, $\langle\delta_{\rm gal}\rangle_{\rm corr} = M_{\rm tot}/(V_{\rm corr} \rho_m)-1$.
These values for each of the ten peaks with \mbox{$3.20<z_{\rm peak}<3.45$} and \mbox{log($M_{\rm tot}$/M$_\odot$)>12} are given in Table~\ref{tab:peaks_id}.

As a cursory test of the effects of targeted followup spectroscopy on the identification and characterization of these systems, we perform the same search and calculations on a VMC map which does not include the C3VO and MAGAZ3NE observations.
This search identifies peaks consistent with six of the ten detected in our fiducial maps (P2, P3, P4, P5, P8, P10) with mass differences from the original characterization ranging from 0.1~dex more massive to 1.0~dex less massive, with a median difference of 0.35~dex less massive.
Based on simulation comparisons \citep[][D. Hung et al., in preparation]{Hung2020} and considering the addition of new data, these differences are consistent with the statistical and systematic errors of mass determination for systems at these epochs of $\sim0.5$~dex.
The peak with the most discrepant masses in the two reconstructuions (P10) lies just on the boundary of the VUDS field of view, while three of the four unmatched peaks (P1, P6, P9) lie outside of the VUDS field of view (Figure~\ref{fig:Structure}) and are thus not expected to be identified due to the significant fraction of confirmed galaxies at this epoch which come from this survey (center panel of Figure~\ref{fig:Specz}).
The final unmatched peak (P7) appears as an extended region of $\sim2\sigma$ overdensity, and thus does not meet the $5\sigma$ peak criterion.
We conclude that the exclusion of targeted spectroscopic observations would not change the fact that we have identified a large number of overdense peaks in a relatively small volume, and would only have a modest effect on the derived richness of the structures.

\subsection{Characterization of the Larger Structure(s)}

As seen in previous studies of other high-redshift protoclusters, \eg\ the Hyperion proto-supercluster in COSMOS at $z\sim2.45$ \citep{Cucciati2018} and the $z\sim3.3$ structure in the XMM-LSS field \citep{Shen2021}, multiple peaks may be embedded in a single more extended plateau of lower significance which is still statistically above the field.
Indeed, we see several peaks in close proximity to each other that may be connected by lower density bridges.
Similar to the aforementioned studies, we perform a search for overdensities but instead use an overdensity significance contour of $\sigma_\delta>2$ and the center of each previously determined peak as a seed.

This results in six extended overdense structures, one of which contains 5 overdense peaks encapsulated within a $\sigma_\delta=2$ contour, and each of the remaining five structures containing a single peak.
We calculate similar values as above for each of these extended structures, listed in Table~\ref{tab:struct_id}, and show the location of both the structures and peaks in Figure~\ref{fig:Structure}.
However we do not perform the same correction for elongation as these structures appear more extended and non-spherical in shape in the VMC maps.

\begin{table*}
	\caption{The properties of $>2\sigma$ extended overdense structures containing the peaks listed in Table~\ref{tab:peaks_id}. In addition to properties similar to Table~\ref{tab:peaks_id}, the number of spectroscopically confirmed members (5), the spectroscopic redshift fraction (the fraction of photometric galaxies within the projected outline of the structure with $3.20<z_{\rm phot}<3.45$ which have a high quality spectroscopic redshift; 6), and the peaks lying within the structure (10) are listed.}
	\begin{center}
	\begin{threeparttable}
	\begin{tabular}{cccccccccc}
		\hline
		ID	& R.A.$_{\rm str}$  & Dec.$_{\rm str}$  & $z_{\rm str}$  & $n_{\rm spec}$  & SzF  &  $\langle \delta_{\rm gal} \rangle$  & V	& $M_{\rm tot}$	 & Peaks\\
			& (deg)		    & (deg)			 &			 &			      & 	  &			       & (cMpc$^3$)	& ($10^{14}M_\odot$) & \\
		(1) & (2) & (3) & (4)  & (5)  & (6) & (7) & (8) & (9) & (10) \\
\hline
                 S1 & 149.8718 & 2.3519 & 3.324 & 80 & 0.295 & 1.09 & 24538	& 13.54 & P1, P2, P3, P6, P8\\
                 S2 & 150.4053 & 2.3386 & 3.260 & 7   & 0.079 & 0.96 & 2546	& 1.36 & P4\\
                 S3 & 150.2166 & 2.4735 & 3.341 & 18 & 0.122 & 0.94 & 5297	& 2.82 & P5\\
                 S4 & 150.1122 & 2.5503 & 3.354 & 9   & 0.421 & 1.08 & 1715	& 0.94 & P7\\
                 S5 & 149.7422 & 2.5856 & 3.410 & 5   & 0.227 & 0.98 & 2098	& 1.13 & P9\\
                 S6 & 150.4499 & 2.5611 & 3.334 & 20 & 0.020 & 0.77 & 177950\tnote{1} & 90.45\tnote{1} & P10 \\
	\end{tabular}
	\begin{tablenotes}
	\item[1] Due to the unconstrained projected extent and low SzF of S6, the derived volume and mass are highly uncertain.
	\end{tablenotes}
	\end{threeparttable}
	\end{center}
	\label{tab:struct_id}
\end{table*}

\section{Evolution of the Overdense Peaks}\label{Sec:Evo}

\begin{figure}
	\includegraphics[width=0.5\textwidth]{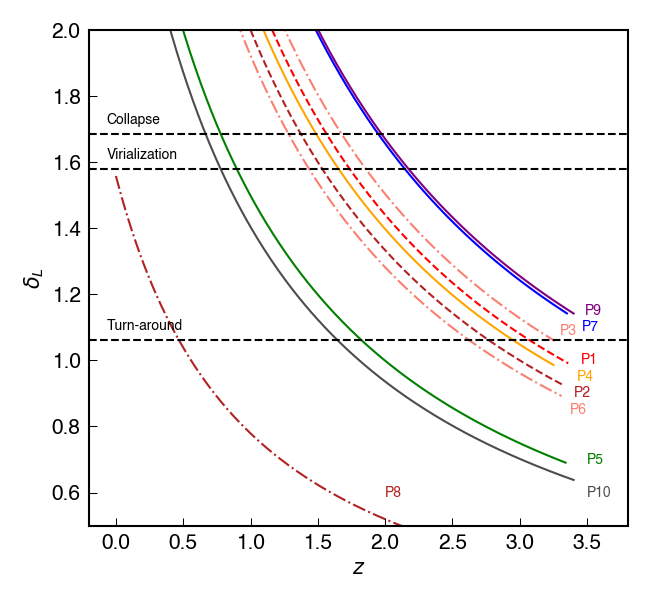}
    \caption{Predicted dynamical evolution of the 10 overdense peaks of \PSC\ as a function of redshift assuming linear spherical collapse. Each peak is identified by number in the plot and colored similarly to its parent structure in Figure~\ref{fig:Structure}. The multiple peaks embedded in S1 (shades of red) are dashed / dotted lines. In isolation, the majority of the peaks will virialize by $z\sim1.5$, though interactions with other nearby peaks make this a rough estimate for the peaks in S1.}
    \label{fig:Evolution}
\end{figure}

As in \citet{Cucciati2018} and \citet{Shen2021}, we characterize the collapse of the overdense peaks assuming spherical linear collapse, which has been shown to give a rough approximation of the dynamical evolution of similar systems \citep[\eg][]{Despali2013}.
Elongation-corrected overdensities are converted to mass overdensities via a bias factor of 3.1 \citep{Durkalec2015} and are subsequently transformed to a linear density, $\delta_{L}$, as in \citet{Bernardeau1994, Steidel1998}.
The results of these calculations are shown in Figure~\ref{fig:Evolution}.

When a region exceeds a critical overdensity, the system will eventually separate from the Hubble flow as self-gravitation slows its expansion to a halt and it attains some maximal physical size at ``turn-around" ($\delta_{L,ta}\sim1.06$), after which the gravitationally-bound structure begins to collapse.
Within this framework, the structure is then virialized once the radius of the structure has contracted to half of its value at turn-around ($\delta_{L,v}\sim1.58$) and, in principle, eventually reaches a final collapsed state ($\delta_{L,c}\sim1.69$).

We model the evolution of each peak by treating each individually, and find that two peaks are predicted to enter their collapse phase by $z\sim2$, while five others will do so by $z\sim1.3$.
This makes the overarching system an ideal ground for studying the diversity of coeval overdense peaks in different stages of evolution, as well as analyzing the properties of their constituent galaxies.

The above calculation assumes independent evolution of the peaks, an assumption that is more reasonable for some peaks than others.
In particular, Peak 1 and Peak 6 are separated in projected space by $<1$~pMpc and have a velocity difference of $\Delta v \sim 3500$~km/s, indicating they are extremely likely to interact in the $\sim1.4$~Gyr before Peak 1 is modeled to virialize.
Similarly, Peaks 2, 3, and 8 are separated in projected space by $<2$~pMpc and $\Delta v \sim 6000$~km/s which could also result in interactions between these peaks.
The remaining peaks are at projected distances $d>3$~pMpc from each other which would require $z\sim0$ halo masses $M_h\gtrsim 2\times10^{15}$~M$_\odot$ to merge together by the present day \citep{Chiang2013, Muldrew2015}.
While this total mass may well be contained within the extended structures presented herein, it likely is not within the individual peaks, suggesting their evolution is unlikely to be affected by other peaks before the modeled epoch of collapse at $z\sim1-2.5$, the above exceptions notwithstanding.

\section{Conclusions}\label{Sec:Conc}

We have presented ten $>5\sigma$ overdense peaks at $3.20<z<3.45$ in the COSMOS field.
This remarkable number of overdense peaks at a similar epoch in a relatively small field of view supports the hypothesis that these structures are all parts of some larger protostructure.
Protoclusters at these epochs are predicted by simulations to have large radii \citep[\eg\ $R_e(\log(M/$M$_\odot)\sim15)$/cMpc $\sim 10$;][]{Chiang2013} and the two most separated peaks herein (P1, P10) have a projected comoving separation of $\sim 70$~cMpc which suggests the presence of a larger structure.
The Hyperion proto-supercluster at $z\sim2.45$ spans a volume of $60\times60\times150$ cMpc$^3$, which is similar in size to \PSC\ and indicates the two structures may be similar in nature. 

The structure with the most confirmed members (S1) in this work has a total mass of \mbox{$M_{\rm tot}=1.35\times 10^{15}~$M$_\odot$} and has 80 spectroscopically confirmed members with redshifts in the range $3.25<z<3.38$.
It also contains five overdense peaks (P1, P2, P3, P6, P8), three of which have masses \mbox{$M_{\rm tot}>2\times 10^{13}~$M$_\odot$} and average overdensities $\langle \delta_{\rm gal, corr} \rangle \gtrsim 10$.
Five other structures contain a single peak (S2-P4, S3-P5, S4-P7, S5-P9, S6-P10) and have peak redshifts in the range $3.25<z<3.43$. Four of these have total masses of \mbox{$M_{\rm tot}\sim1\times 10^{14}~$M$_\odot$}, while the extent and mass of S6 are poorly constrained due to the structure's location on the edge of the region with significant spectroscopic coverage.

From the VMC maps alone, the final structure (S6-P10) has a much larger projected extent and redshift depth with a lower average peak overdensity $\langle \delta_{\rm gal, corr} \rangle \sim 4$.
Much of the structure also falls outside the field covered by the VUDS survey (and has not been targeted by C3VO or MAGAZ3NE), resulting in fewer spectroscopically confirmed galaxies (spectroscopically-confirmed fraction $<2\%$) and therefore a smoothing out of the structure in the redshift dimension.
As a result, the uncertainties on the true extent of this structure and its mass are significant and it is of particular interest for spectroscopic followup to confirm the extent of the overdensity.

The ten peaks spanning almost two orders of magnitude in mass at the same redshift within a larger system make \PSC\ an excellent target for probing the effects of environment on galaxy evolution at these early times, as will be done in upcoming work.

\begin{table}
	\centering
	\caption{The evolution of the 10 overdense peaks (1) assuming linear spherical collapse. Included are the peak redshift (2), bias-corrected matter overdensity (3), the modeled redshifts of turn-around, when the structure reaches its greatest physical extent ($z_{\rm ta}$, 4) and collapse, and $z_{\rm c}$, 5), and the time between these redshifts and $z_{\rm peak}$ (6-7).}
	\label{tab:evo}
	\begin{tabular}{ccccccc}
		\hline
		ID  & $z_{\rm peak}$  & $\langle \delta_{\rm m} \rangle$  & $z_{\rm ta}$  & $z_{\rm c}$  & $\Delta t_{\rm ta}$  & $\Delta t_{\rm c}$  \\
		      &		    		 &						     &				&		  & (Gyr)			    & 	(Gyr)	 \\
		(1)  & (2) 			    & (3) 					& (4)  			 & (5)  	& (6) 		    & (7) 	\\
		\hline
                 P1 & 3.366 & 4.03 & 3.07 & 1.56 & 0.21 & 2.21 \\
                 P2 & 3.340 & 3.18 & 2.77 & 1.37 & 0.43 & 2.66 \\
                 P3 & 3.269 & 5.22 & 3.25 & 1.68 & 0.01 & 1.88 \\
                 P4 & 3.255 & 3.99 & 2.95 & 1.49 & 0.22 & 2.29 \\
                 P5 & 3.343 & 1.52 & 1.82 & 0.78 & 1.66 & 4.87 \\
                 P6 & 3.315 & 2.88 & 2.62 & 1.28 & 0.56 & 2.90 \\
                 P7 & 3.354 & 7.59 & $>z_{\rm obs}$ & 1.95 & --      & 1.44 \\
                 P8 & 3.248 & 0.52 & 0.47 & Future & 6.71 & -- \\
                 P9 & 3.408 & 7.14 & $>z_{\rm obs} $ & 1.95 & --      & 1.48 \\
                 P10 & 3.409 & 1.29 & 1.64 & 0.67 & 2.06 & 5.51 \\
		\hline
	\end{tabular}
\end{table}

\section*{Acknowledgements}

Some of the data presented herein were obtained at the W. M. Keck Observatory, which is operated as a scientific partnership among the California Institute of Technology, the University of California and the National Aeronautics and Space Administration. The Observatory was made possible by the generous financial support of the W. M. Keck Foundation.
The authors wish to recognize and acknowledge the very significant cultural role and reverence that the summit of Maunakea has always had within the indigenous Hawaiian community.  We are most fortunate to have the opportunity to conduct observations from this mountain.
This material is based upon work supported by the National Science Foundation under Grant No. 1908422.
GG and MT acknowledge the support from grant PRIN MIUR 20173ML3WW\_001.
GW gratefully acknowledges support from the National Science Foundation through grant AST-2205189 and from HST program numbers GO-15294 and GO-16300.
Support for program numbers GO-15294 and GO-16300 was provided by NASA through grants from the Space Telescope Science Institute, which is operated by the Association of Universities for Research in Astronomy, Incorporated, under NASA contract NAS5-26555.
DM acknowledges support by the National Science Foundation Grant AST-2009442 and by the National Aeronautics and Space Administration (NASA) under award number 80NSSC21K0630, issued through the
NNH20ZDA001N Astrophysics Data Analysis Program (ADAP).

\section*{Data Availability}

 The data underlying this article will be shared on reasonable request to the corresponding author.



\bibliographystyle{mnras}
\bibliography{library} 

\begin{thebibliography}{}
\makeatletter
\relax
\def\mn@urlcharsother{\let\do\@makeother \do\$\do\&\do\#\do\^\do\_\do\%\do\~}
\def\mn@doi{\begingroup\mn@urlcharsother \@ifnextchar [ {\mn@doi@}
  {\mn@doi@[]}}
\def\mn@doi@[#1]#2{\def\@tempa{#1}\ifx\@tempa\@empty \href
  {http://dx.doi.org/#2} {doi:#2}\else \href {http://dx.doi.org/#2} {#1}\fi
  \endgroup}
\def\mn@eprint#1#2{\mn@eprint@#1:#2::\@nil}
\def\mn@eprint@arXiv#1{\href {http://arxiv.org/abs/#1} {{\tt arXiv:#1}}}
\def\mn@eprint@dblp#1{\href {http://dblp.uni-trier.de/rec/bibtex/#1.xml}
  {dblp:#1}}
\def\mn@eprint@#1:#2:#3:#4\@nil{\def\@tempa {#1}\def\@tempb {#2}\def\@tempc
  {#3}\ifx \@tempc \@empty \let \@tempc \@tempb \let \@tempb \@tempa \fi \ifx
  \@tempb \@empty \def\@tempb {arXiv}\fi \@ifundefined
  {mn@eprint@\@tempb}{\@tempb:\@tempc}{\expandafter \expandafter \csname
  mn@eprint@\@tempb\endcsname \expandafter{\@tempc}}}

\bibitem[\protect\citeauthoryear{Arnouts, Cristiani, Moscardini, Matarrese,
  Lucchin, Fontana  \& Giallongo}{Arnouts et~al.}{1999}]{Arnouts1999}
Arnouts S.,  Cristiani S.,  Moscardini L.,  Matarrese S.,  Lucchin F.,  Fontana
  A.,   Giallongo E.,  1999, \mn@doi [Monthly Notices of the Royal Astronomical
  Society] {10.1046/j.1365-8711.1999.02978.x}, 310, 540

\bibitem[\protect\citeauthoryear{Balogh et~al.,}{Balogh
  et~al.}{2020}]{Balogh2020}
Balogh M.~L.,  et~al., 2020, \mn@doi [Monthly Notices of the Royal Astronomical
  Society] {10.1093/mnras/staa3008}, 31, 1

\bibitem[\protect\citeauthoryear{Bernardeau}{Bernardeau}{1994}]{Bernardeau1994}
Bernardeau F.,  1994, \mn@doi [The Astrophysical Journal] {10.1086/174121},
  427, 51

\bibitem[\protect\citeauthoryear{Capak et~al.,}{Capak et~al.}{2007}]{Capak2007}
Capak P.,  et~al., 2007, \mn@doi [The Astrophysical Journal Supplement Series]
  {10.1086/519081}, 172, 99

\bibitem[\protect\citeauthoryear{Capak et~al.,}{Capak et~al.}{2011}]{Capak2011}
Capak P.~L.,  et~al., 2011, eprint arXiv:1101.3586, pp 1--16

\bibitem[\protect\citeauthoryear{Casey et~al.,}{Casey et~al.}{2022}]{Casey2022}
Casey C.~M.,  et~al., 2022

\bibitem[\protect\citeauthoryear{Chabrier}{Chabrier}{2003}]{Chabrier2003}
Chabrier G.,  2003, \mn@doi [Publications of the Astronomical Society of the
  Pacific] {10.1086/376392}, 115, 763

\bibitem[\protect\citeauthoryear{Chiang, Overzier  \& Gebhardt}{Chiang
  et~al.}{2013}]{Chiang2013}
Chiang Y.-K.,  Overzier R.,   Gebhardt K.,  2013, \mn@doi [The Astrophysical
  Journal] {10.1088/0004-637X/779/2/127}, 779, 127

\bibitem[\protect\citeauthoryear{Chon, B{\"{o}}hringer  \& Zaroubi}{Chon
  et~al.}{2015}]{Chon2015}
Chon G.,  B{\"{o}}hringer H.,   Zaroubi S.,  2015, \mn@doi [Astronomy and
  Astrophysics] {10.1051/0004-6361/201425591}, 575, 1

\bibitem[\protect\citeauthoryear{Cucciati et~al.,}{Cucciati
  et~al.}{2018}]{Cucciati2018}
Cucciati O.,  et~al., 2018, \mn@doi [Astronomy {\&} Astrophysics]
  {10.1051/0004-6361/201833655}, 619, A49

\bibitem[\protect\citeauthoryear{Despali, Tormen  \& Sheth}{Despali
  et~al.}{2013}]{Despali2013}
Despali G.,  Tormen G.,   Sheth R.~K.,  2013, \mn@doi [Monthly Notices of the
  Royal Astronomical Society] {10.1093/mnras/stt235}, 431, 1143

\bibitem[\protect\citeauthoryear{Durkalec et~al.,}{Durkalec
  et~al.}{2015}]{Durkalec2015}
Durkalec A.,  et~al., 2015, \mn@doi [Astronomy and Astrophysics]
  {10.1051/0004-6361/201425343}, 583, 1

\bibitem[\protect\citeauthoryear{Einasto et~al.,}{Einasto
  et~al.}{2021}]{Einasto2021}
Einasto M.,  et~al., 2021, \mn@doi [Astronomy and Astrophysics]
  {10.1051/0004-6361/202040200}, 649, 1

\bibitem[\protect\citeauthoryear{Faber et~al.,}{Faber et~al.}{2003}]{Faber2003}
Faber S.~M.,  et~al., 2003, \mn@doi [Instrument Design and Performance for
  Optical/Infrared Ground-based Telescopes] {10.1117/12.460346}, 4841, 1657

\bibitem[\protect\citeauthoryear{Forrest et~al.,}{Forrest
  et~al.}{2017}]{Forrest2017}
Forrest B.,  et~al., 2017, \mn@doi [The Astrophysical Journal]
  {10.3847/2041-8213/aa653b}, 838, L12

\bibitem[\protect\citeauthoryear{Forrest et~al.,}{Forrest
  et~al.}{2020}]{Forrest2020b}
Forrest B.,  et~al., 2020, \mn@doi [The Astrophysical Journal]
  {10.3847/1538-4357/abb819}, 903, 47

\bibitem[\protect\citeauthoryear{Hasinger et~al.,}{Hasinger
  et~al.}{2018}]{Hasinger2018}
Hasinger G.,  et~al., 2018, \mn@doi [The Astrophysical Journal]
  {10.3847/1538-4357/aabacf}, 858, 77

\bibitem[\protect\citeauthoryear{Horne}{Horne}{1986}]{Horne1986}
Horne K.,  1986, \mn@doi [Publications of the Astronomical Society of the
  Pacific] {10.1086/131801}, 98, 609

\bibitem[\protect\citeauthoryear{Hung et~al.,}{Hung et~al.}{2020}]{Hung2020}
Hung D.,  et~al., 2020, \mn@doi [Monthly Notices of the Royal Astronomical
  Society] {10.1093/mnras/stz3164}, 491, 5524

\bibitem[\protect\citeauthoryear{Hung et~al.,}{Hung et~al.}{2021}]{Hung2021}
Hung D.,  et~al., 2021, \mn@doi [Monthly Notices of the Royal Astronomical
  Society] {10.1093/mnras/stab300}, 502, 3942

\bibitem[\protect\citeauthoryear{Ilbert et~al.,}{Ilbert
  et~al.}{2006}]{Ilbert2006}
Ilbert O.,  et~al., 2006, \mn@doi [Astronomy and Astrophysics]
  {10.1051/0004-6361:20065138}, 457, 841

\bibitem[\protect\citeauthoryear{Ilbert et~al.,}{Ilbert
  et~al.}{2009}]{Ilbert2009}
Ilbert O.,  et~al., 2009, \mn@doi [The Astrophysical Journal]
  {10.1088/0004-637X/690/2/1236}, 690, 1236

\bibitem[\protect\citeauthoryear{Ito et~al.,}{Ito et~al.}{2023}]{Ito2023}
Ito K.,  et~al., 2023, \mn@doi [The Astrophysical Journal Letters]
  {10.3847/2041-8213/acb49b}, 945, L9

\bibitem[\protect\citeauthoryear{Jin et~al.,}{Jin et~al.}{2021}]{Jin2021}
Jin S.,  et~al., 2021, \mn@doi [Astronomy and Astrophysics]
  {10.1051/0004-6361/202040232}, 652, 1

\bibitem[\protect\citeauthoryear{Koekemoer et~al.,}{Koekemoer
  et~al.}{2007}]{Koekemoer2007}
Koekemoer A.~M.,  et~al., 2007, \mn@doi [The Astrophysical Journal Supplement
  Series] {10.1086/520086}, 172, 196

\bibitem[\protect\citeauthoryear{Kriek et~al.,}{Kriek et~al.}{2015}]{Kriek2015}
Kriek M.,  et~al., 2015, \mn@doi [The Astrophysical Journal Supplement Series]
  {10.1088/0067-0049/218/2/15}, 218, 1

\bibitem[\protect\citeauthoryear{Laigle et~al.,}{Laigle
  et~al.}{2016}]{Laigle2016}
Laigle C.,  et~al., 2016, \mn@doi [The Astrophysical Journal Supplement Series]
  {10.3847/0067-0049/224/2/24}, 224, 24

\bibitem[\protect\citeauthoryear{{Le F{\`{e}}vre} et~al.,}{{Le F{\`{e}}vre}
  et~al.}{2003}]{Lefevre2003}
{Le F{\`{e}}vre} O.,  et~al., 2003, \mn@doi [Instrument Design and Performance
  for Optical/Infrared Ground-based Telescopes] {10.1117/12.460959}, 4841, 1670

\bibitem[\protect\citeauthoryear{{Le F{\`{e}}vre} et~al.,}{{Le F{\`{e}}vre}
  et~al.}{2015}]{LeFevre2015}
{Le F{\`{e}}vre} O.,  et~al., 2015, \mn@doi [Astronomy {\&} Astrophysics]
  {10.1051/0004-6361/201423829}, 576, A79

\bibitem[\protect\citeauthoryear{Lemaux et~al.,}{Lemaux
  et~al.}{2014}]{Lemaux2014}
Lemaux B.~C.,  et~al., 2014, \mn@doi [Astronomy and Astrophysics]
  {10.1051/0004-6361/201423828}, 572, 1

\bibitem[\protect\citeauthoryear{Lemaux, Tomczak, Lubin, Wu, Gal, Rumbaugh,
  Kocevski  \& Squires}{Lemaux et~al.}{2017}]{Lemaux2017}
Lemaux B.~C.,  Tomczak A.~R.,  Lubin L.~M.,  Wu P.-F.,  Gal R.~R.,  Rumbaugh
  N.,  Kocevski D.~D.,   Squires G.~K.,  2017, \mn@doi [Monthly Notices of the
  Royal Astronomical Society] {10.1093/mnras/stx1579}, 472, 419

\bibitem[\protect\citeauthoryear{Lemaux et~al.,}{Lemaux
  et~al.}{2018}]{Lemaux2018}
Lemaux B.~C.,  et~al., 2018, \mn@doi [Astronomy {\&} Astrophysics]
  {10.1051/0004-6361/201730870}, 615, A77

\bibitem[\protect\citeauthoryear{Lemaux et~al.,}{Lemaux
  et~al.}{2022}]{Lemaux2022}
Lemaux B.~C.,  et~al., 2022, \mn@doi [Astronomy {\&} Astrophysics]
  {10.1051/0004-6361/202039346}, 662, A33

\bibitem[\protect\citeauthoryear{Lilly et~al.,}{Lilly et~al.}{2007}]{Lilly2007}
Lilly S.~J.,  et~al., 2007, \mn@doi [The Astrophysical Journal Supplement
  Series] {10.1086/516589}, 172, 70

\bibitem[\protect\citeauthoryear{McConachie et~al.,}{McConachie
  et~al.}{2022}]{McConachie2022}
McConachie I.,  et~al., 2022, \mn@doi [The Astrophysical Journal]
  {10.3847/1538-4357/ac2b9f}, 926, 37

\bibitem[\protect\citeauthoryear{McLean et~al.,}{McLean
  et~al.}{2010}]{McLean2010}
McLean I.~S.,  et~al., 2010, in McLean I.~S.,  Ramsay S.~K.,   Takami H.,  eds,
   Vol. 7735, Ground-based and Airborne Instrumentation for Astronomy III. p.
  77351E, \mn@doi{10.1117/12.856715}, \url
  {http://proceedings.spiedigitallibrary.org/proceeding.aspx?doi=10.1117/12.856715}

\bibitem[\protect\citeauthoryear{McLean et~al.,}{McLean
  et~al.}{2012}]{McLean2012}
McLean I.~S.,  et~al., 2012, in McLean I.~S.,  Ramsay S.~K.,   Takami H.,  eds,
   Vol. 8446, Ground-based and Airborne Instrumentation for Astronomy IV. ,
  \mn@doi{10.1117/12.924794}, \url
  {http://proceedings.spiedigitallibrary.org/proceeding.aspx?doi=10.1117/12.924794}

\bibitem[\protect\citeauthoryear{Muldrew, Hatch  \& Cooke}{Muldrew
  et~al.}{2015}]{Muldrew2015}
Muldrew S.~I.,  Hatch N.~A.,   Cooke E.~A.,  2015, \mn@doi [Monthly Notices of
  the Royal Astronomical Society] {10.1093/mnras/stv1449}, 452, 2528

\bibitem[\protect\citeauthoryear{Muzzin et~al.,}{Muzzin
  et~al.}{2013}]{Muzzin2013a}
Muzzin A.,  et~al., 2013, \mn@doi [The Astrophysical Journal Supplement Series]
  {10.1088/0067-0049/206/1/8}, 206, 8

\bibitem[\protect\citeauthoryear{Nayyeri et~al.,}{Nayyeri
  et~al.}{2017}]{Nayyeri2017}
Nayyeri H.,  et~al., 2017, \mn@doi [The Astrophysical Journal Supplement
  Series] {10.3847/1538-4365/228/1/7}, 228, 7

\bibitem[\protect\citeauthoryear{Oke \& Gunn}{Oke \& Gunn}{1983}]{Oke1983}
Oke J.~B.,  Gunn J.~E.,  1983, \mn@doi [The Astrophysical Journal]
  {10.1086/160817}, 266, 713

\bibitem[\protect\citeauthoryear{Ouchi et~al.,}{Ouchi et~al.}{2005}]{Ouchi2005}
Ouchi M.,  et~al., 2005, \mn@doi [The Astrophysical Journal] {10.1086/428499},
  620, L1

\bibitem[\protect\citeauthoryear{Pentericci et~al.,}{Pentericci
  et~al.}{2000}]{Pentericci2000}
Pentericci L.,  et~al., 2000, Astronomy and Astrophysics, 361, 5

\bibitem[\protect\citeauthoryear{Raychaudhury, Fabian, Edge, Jones  \&
  Forman}{Raychaudhury et~al.}{1991}]{Raychaudhury1991}
Raychaudhury S.,  Fabian A.~C.,  Edge A.~C.,  Jones C.,   Forman W.,  1991,
  \mn@doi [Monthly Notices of the Royal Astronomical Society]
  {10.1093/mnras/248.1.101}, 248, 101

\bibitem[\protect\citeauthoryear{Remus, Dolag  \& Dannerbauer}{Remus
  et~al.}{2023}]{Remus2023}
Remus R.-S.,  Dolag K.,   Dannerbauer H.,  2023, \mn@doi [The Astrophysical
  Journal] {10.3847/1538-4357/accb91}, 950, 191

\bibitem[\protect\citeauthoryear{Scoville et~al.,}{Scoville
  et~al.}{2007}]{Scoville2007}
Scoville N.,  et~al., 2007, \mn@doi [The Astrophysical Journal Supplement
  Series] {10.1086/516585}, 172, 1

\bibitem[\protect\citeauthoryear{Shen et~al.,}{Shen et~al.}{2021}]{Shen2021}
Shen L.,  et~al., 2021, \mn@doi [The Astrophysical Journal]
  {10.3847/1538-4357/abee75}, 912, 60

\bibitem[\protect\citeauthoryear{Steidel, Adelberger, Dickinson, Giavalisco,
  Pettini  \& Kellogg}{Steidel et~al.}{1998}]{Steidel1998}
Steidel C.~C.,  Adelberger K.~L.,  Dickinson M.,  Giavalisco M.,  Pettini M.,
  Kellogg M.,  1998, \mn@doi [The Astrophysical Journal] {10.1086/305073}, 492,
  428

\bibitem[\protect\citeauthoryear{Tomczak et~al.,}{Tomczak
  et~al.}{2017}]{Tomczak2017}
Tomczak A.~R.,  et~al., 2017, \mn@doi [Monthly Notices of the Royal
  Astronomical Society] {10.1093/mnras/stx2245}, 472, 3512

\bibitem[\protect\citeauthoryear{Tully, Courtois, Hoffman  \&
  Pomar{\`{e}}de}{Tully et~al.}{2014}]{Tully2014}
Tully R.~B.,  Courtois H.,  Hoffman Y.,   Pomar{\`{e}}de D.,  2014, \mn@doi
  [Nature] {10.1038/nature13674}, 513, 71

\bibitem[\protect\citeauthoryear{Venemans et~al.,}{Venemans
  et~al.}{2002}]{Venemans2002}
Venemans B.~P.,  et~al., 2002, \mn@doi [The Astrophysical Journal]
  {10.1086/340563}, 569, L11

\bibitem[\protect\citeauthoryear{Weaver et~al.,}{Weaver
  et~al.}{2022}]{Weaver2022a}
Weaver J.~R.,  et~al., 2022, \mn@doi [The Astrophysical Journal Supplement
  Series] {10.3847/1538-4365/ac3078}, 258, 11

\bibitem[\protect\citeauthoryear{Yamada, Nakamura, Matsuda, Hayashino,
  Yamauchi, Morimoto, Kousai  \& Umemura}{Yamada et~al.}{2012}]{Yamada2012}
Yamada T.,  Nakamura Y.,  Matsuda Y.,  Hayashino T.,  Yamauchi R.,  Morimoto
  N.,  Kousai K.,   Umemura M.,  2012, \mn@doi [Astronomical Journal]
  {10.1088/0004-6256/143/4/79}, 143

\bibitem[\protect\citeauthoryear{van~der Burg, Muzzin, Hoekstra, Wilson, Lidman
   \& Yee}{van~der Burg et~al.}{2014}]{vanderBurg2014}
van~der Burg R. F.~J.,  Muzzin A.,  Hoekstra H.,  Wilson G.,  Lidman C.,   Yee
  H. K.~C.,  2014, \mn@doi [Astronomy {\&} Astrophysics]
  {10.1051/0004-6361/201322771}, 561, A79

\makeatother
\end{thebibliography}






\bsp	
\label{lastpage}
\end{document}